\documentclass[aps,prl,twocolumn,showpacs,floatfix,superscriptaddress]{revtex4}

\usepackage{amsmath,amssymb}
\usepackage{graphicx}
\usepackage{times}

\newcommand{\varH}{{\mathcal{H}}}
\newcommand{\varI}{{\mathcal{I}}}
\newcommand{\varN}{{\mathcal{N}}}
\newcommand{\varO}{{\mathcal{O}}}

\newcommand{\bfk}{{\mathbf{k}}}

\let\up\uparrow
\let\down\downarrow

\newcommand{\comm}[2]{[#1,#2]}
\newcommand{\Braket}[1]{\mathinner{\langle{\textstyle#1}\rangle}}

\newcommand{\im}{{\rm Im}}
\newcommand{\re}{{\rm Re}}

\newcommand{\eqnref}[1]{Eq.~(\ref{#1})}
\newcommand{\Figref}[1]{Figure~\ref{#1}}
\newcommand{\figref}[1]{Fig.~\ref{#1}}
\newcommand{\figsref}[1]{Figs.~\ref{#1}}

\begin{document}
\title{Many-body Correlation Effect on Mesoscopic Charge Relaxation}
\author{Minchul Lee}
\affiliation{Department of Applied Physics, College of Applied Science, Kyung Hee University, Yongin 446-701, Korea}
\author{Rosa L\'opez}
\affiliation{Departament de F\'{i}sica, Universitat de les Illes Balears,
  E-07122 Palma de Mallorca, Spain}
\affiliation{Institut de F\'{i}sica Interdisciplinar i de Sistemes Complexos
  IFISC (CSIC-UIB), E-07122 Palma de Mallorca, Spain}
\author{Mahn-Soo Choi}
\affiliation{Department of Physics, Korea University, Seoul 136-701, Korea}
\author{Thibaut Jonckheere}
\affiliation{Centre de Physique Th\'eorique, UMR6207, Case 907, Luminy, 13288 Marseille Cedex 9, France}
\author{Thierry Martin}
\affiliation{Centre de Physique Th\'eorique, UMR6207, Case 907, Luminy, 13288 Marseille Cedex 9, France}
\affiliation{Universit\'e de la M\'editerran\'ee, 13288 Marseille Cedex 9, France}

\begin{abstract}
  We investigate in a nonperturbative way the dynamics of a correlated quantum
  capacitor. We find that the many-body correlations do not disturb the
  universal low-frequency relaxation resistance per channel, $R_q(\omega=0) =
  h/4e^2$ ensured by the Korringa-Shiba rule whereas the interpretation of the
  quantum capacitance $C_q$ in terms of the density of states fails when strong
  correlations are present.  The AC resistance $R_q(\omega)$ shows huge peaks
  (with values larger than $h/4e^2$) at $\hbar\omega \approx \pm \Gamma^*$,
  where $\Gamma^*$ is the renormalized level broadening. These peaks are
  merged to a single one at $\omega=0$ when a finite Zeeman field is applied
  comparable to $\Gamma^*$. The observed features of $R_q$, being most evident
  in the Kondo regime, are attributed to the generation of particle-hole
  excitations in the contacts accomplished by spin-flip processes in the dot.
\end{abstract}

\maketitle

\paragraph{Introduction.---}
Optoelectronic devices such as light sources emitting single photons on demand
are of an enormous interest in quantum information.  Recently, its solid-state
analogue, a \textit{quantum capacitor} (QC), was created by F\`{e}ve \textit{et
  al.} \cite{Feve:07}, in which a quantum dot (QD) was coupled to a single
reservoir via a quantum point contact.  Fast time-controlled variations of
fractions of nanoseconds on the dot gate potential produced the single-electron
source of emitting electrons in a desirable quantum state.  Previously, Gabelli
\textit{et al.} \cite{Gabelli:06} showed that the QC could act as a RC circuit
with a quantized resistance as predicted in a series of seminal works
\cite{Buttiker:93a,Buttiker:93b,Nigg:06}.  The experimental interest in the AC
properties of mesoscopic conductors
\cite{Pieper:80,Kouwenhoven:94,Reznikov:95,Verghese:95,Schoelkopf:97,Reydellet:03}
has been revived recently due to the experimental confirmation of the
quantization of both the AC current \cite{Feve:07} and the quantum resistance
\cite{Gabelli:06}.  In a coherent conductor the AC transport is highly
sensitive to the internal distribution of charges and potentials that need to
be calculated in a self-consistently manner to ensure a gauge invariant and
current conserving description.  For a macroscopic capacitor the low-frequency
dynamical conductance depends on two elements: the geometrical capacitance $C$,
and the resistance $R$.  Classically, the electric field on the surface of the
metallic plates is completely screened and $C$ is characterized solely via
Coulomb forces. However, B\"{u}ttiker \textit{et al.} pointed out that for a
coherent nanoscale system that rule statement is not longer valid, since
electric fields penetrate at distances of the order of the Thomas-Fermi
screening length \cite{Buttiker:93a,Buttiker:93b, Buttiker:93c,
  Petre:96}. Consequently, the capacitance for a coherent conductor, termed as
\textit{electrochemical capacitance}, depends on the geometry, on its physical
properties and particularly on its density of states (DOS) through the quantum
capacitance $C_q$. Even more surprisingly, the resistance becomes quantized
independently of the value of the transmission through the mesoscopic conductor
\cite{Landauer:70,Buttiker:86,Landauer:87,Imry:99}. To distinguish it from the
DC resistance, $R_q$ is called \textit{charge relaxation resistance}, and
together with $C_\mu=(C^{-1}+C_q^{-1})^{-1}$ defines the RC time for a QC, that
is, the charge relaxation time upon the action of time-dependent potential.
\begin{figure}[!b]
  \centering
  \includegraphics[width=85mm]{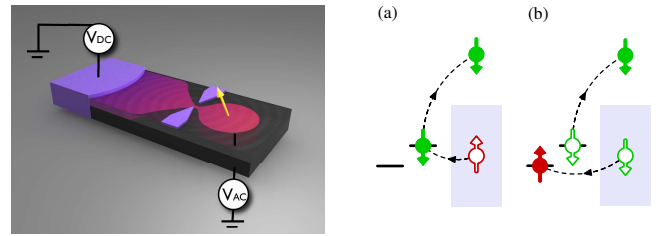}
  \caption{(color online) (LEFT) Illustration of a quantum capacitor and
    (RIGHT) second-order tunneling processes that generate a single
    particle-hole pair in the conduction band without [(a)] and with [(b)] a
    spin flip in the dot. Filled and empty arrows in the contact indicates
    particles and holes, respectively. Here we assume $\Delta_Z>0$.}
  \label{fig:1}
\end{figure}

So far several aspects of AC conductance for RC quantum circuits have been
addressed
\cite{Mora:010,Karyn:02,Ringel:08,Mateev:95,Petitjean:09,Splettstoesser:010,Hamamoto:010,Moskalets:08}
focusing mostly on the spin polarized case.  Therefore the understanding on the
effect of many-body correlations in RC circuits is still missing. This Letter
attempts to fill this gap, offering a physical picture of the influence of
many-body interactions on the dynamics of QCs.  We consider a QC formed by an
ultrasmall QD and calculate its $C_q$ and $R_q$ by using their relations with
the charge susceptibility in the linear response regime. We perform a thorough
study by tuning the QD from the resonant tunneling regime to the Kondo
regime. In the Kondo regime virtual tunneling transitions between the dot and
the reservoir flip efficiently the dot spin resulting in a many-body singlet
spin state with binding energy given by the Kondo energy $k_B T_K$. First, we
find that the Korringa-Shiba (KS) relation \cite{Shiba:75}, valid only in the
Fermi-liquid regime, ensures the quantization of $R_q$ even in the Kondo regime
as long as the frequency is low enough, $\hbar\omega\ll k_BT_K$. However, it is
pointed out that the relation of $C_q$ with the localized DOS should be revised
when the charge dynamics is frozen due to Kondo correlations. More importantly,
the AC resistance $R_{q}(\omega)$ displays peaks at $\hbar\omega\sim\pm
\Gamma^*$, not present in a mean-field description for the Coulomb interaction.
Here $\Gamma^*$ is the renormalized level broadening of the major tunneling dot
level and $\Gamma^* = k_BT_K$ in the Kondo regime.  Third, a departure from the
universal value for $R_q(\omega=0)$ is achieved when an applied Zeeman field
$\Delta_Z$ is comparable to $\Gamma^*$. The enhancement of $R_q$ at zero and
finite frequencies is attributed to the creation of particle-hole ($p$-$h$)
excitations accompanying a spin-flip in the dot due to the strong Coulomb
interaction, see \figref{fig:1}(b).

\paragraph{Model.---}
The essential features of a mesoscopic RC circuit in the presence of
correlations can be well captured in the Anderson model where an interacting
single-level QD is coupled to a single-channel electron reservoir subject to a
weak time-dependent voltage $V(t) = V_{\rm ac} \cos \omega t$. The Hamiltonian
of this system is $\varH = \varH_{\rm L} + \varH_{\rm D} + \varH_{\rm T}$. The
lead part, $\varH_{\rm L} = \sum_{\bfk\mu} [\epsilon_\bfk + eV(t)]
c_{\bfk\mu}^\dag c_{\bfk\mu}$, describes the noninteracting conduction
electrons with energy $\epsilon_\bfk$ (measured with respect to the Fermi
energy $\epsilon_F = 0$) and spin $\mu$ in the reservoir, and the tunneling of
electrons between the reservoir and the dot is modeled by $\varH_{\rm T} =
\sum_{\bfk\mu} \left[t_\bfk d_\mu^\dag c_{\bfk\mu} + (h.c.)\right]$ in terms of
energy-independent tunneling matrix element, $t_\bfk = t$. The hybridization
between the dot and the lead is characterized by a tunneling amplitude $\Gamma
= \pi \rho_0 |t|^2$ ($\rho_0$ is the contact DOS at the Fermi energy). The dot
Hamiltonian reads $\varH_{\rm D} = \sum_\mu \left[\epsilon_\mu + eV(t)\right]
n_\mu + 2E_{\rm C} n_\up n_\down$, where $n_\mu = d_\mu^\dag d_\mu$ is the dot
occupation operator and $E_{\rm C} = e^2/2C$ is the Coulomb charging
energy. The orbital level $\epsilon_\mu = \epsilon_d - \mu\Delta_Z/2$ is
spin-dependent due to a Zeeman energy $\Delta_Z$.  The time-dependent voltage
$V(t)$ induces the polarization charges $N_U(t)$ between the dot and the gate,
which in turn leads to the time-dependent potential $U(t) = |e|N_U(t)/C$ inside
the dot. Consequently, the applied voltage not only generates a current $I(t)$
between the lead and the dot, but also induces a dot-gate displacement current
$I_d(t) = e (dN_U/dt) = - C (dU/dt)$. Charge conservation requires $I(t) +
I_d(t) = 0$. Assuming that the gate-invariant perturbation $V(t) - U(t)$ is
sufficiently small, the linear response theory leads to the relation,
$I(\omega) = g(\omega) (V(\omega) - U(\omega))$, where $g(t) = (ie/\hbar)
\Braket{\comm{\varI(t)}{\varN}}\Theta(t)$ is the equilibrium correlation
function between the occupation operator $\varN = \sum_\mu n_\mu$ and the
current operator $\varI = e (d\varN/dt)$. Note that the current-density
correlation function $g(\omega)$ is directly related to the charge
susceptibility $\chi_c(t) = -i \Braket{\comm{\varN(t)}{\varN}}\Theta(t)$, which
is preferable for numerical computation, via the relation $g(\omega) = i\omega
(e^2/\hbar) \chi_c(\omega)$. Then, with the help of $I(\omega) = - I_d(\omega)
= -i\omega CU(\omega)$, the dot-lead impedance $Z(\omega) =
V(\omega)/I(\omega)$, which is experimentally accessible, is given by
$Z(\omega) = 1/(-i\omega C) + 1/g(\omega)$. The relaxation resistance and the
quantum capacitance are then expressed in terms of the charge susceptibility as
\begin{equation}
  \label{eq:Rq}
  \frac{R_q(\omega)}{h/e^2}
  \!=\! \re\left[\frac{1}{2\pi i\omega \chi_c(\omega)}\right]\!\!,\
  \frac{e^2/h}{C_q(\omega)}
  \!=\! \im\left[\frac{1}{2\pi i \chi_c(\omega)}\right]\!\!.
\end{equation}
The numerical normalization group (NRG) \cite{Wilson:75,Krishnamurthy:80}
treats the Coulomb interaction in a nonperturbative way being the most adequate
method for computing the charge susceptibility \cite{nrgnote}. While the
imaginary part of the susceptibility is obtained by the NRG procedure its real
part is calculated via the Kramers-Kronig relation. Note that the NRG results
for the finite-frequency linear response in the Kondo regime are known to be
reliable as long as the perturbation is weak enough \cite{Sindel:05}. We focus
on the zero-temperature case and use the contact bandwidth $D$ as the energy
unit. We set $k_B=1$ hereafter.
\begin{figure}[!t]
  \centering
  \includegraphics[width=78mm]{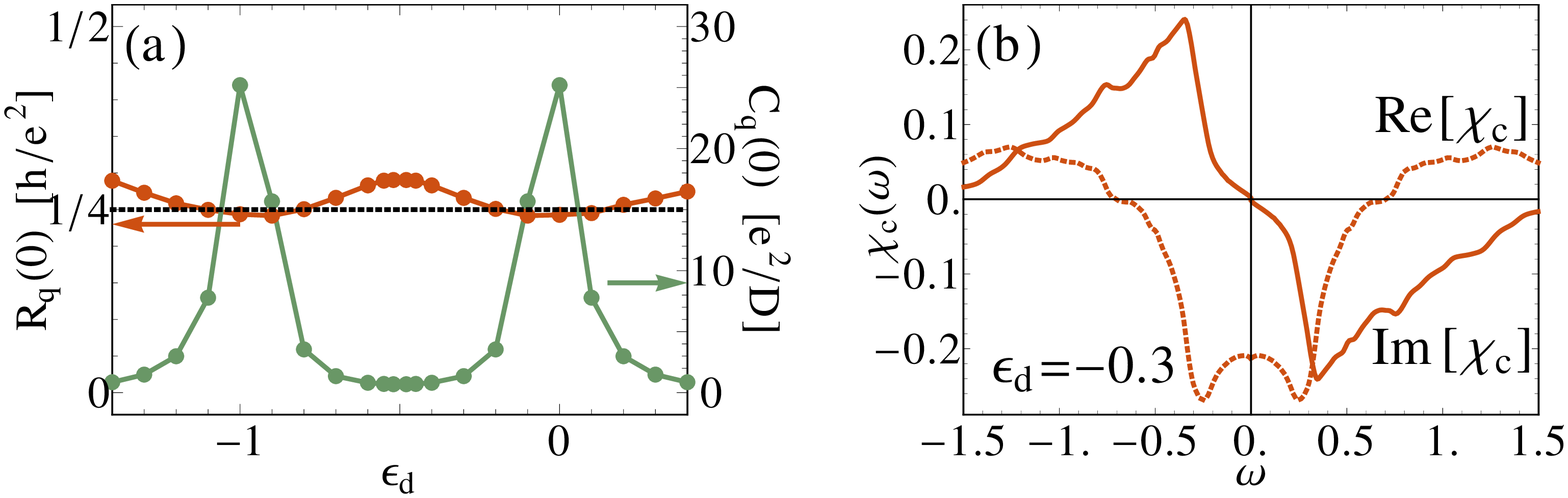}\\
  \includegraphics[width=35mm]{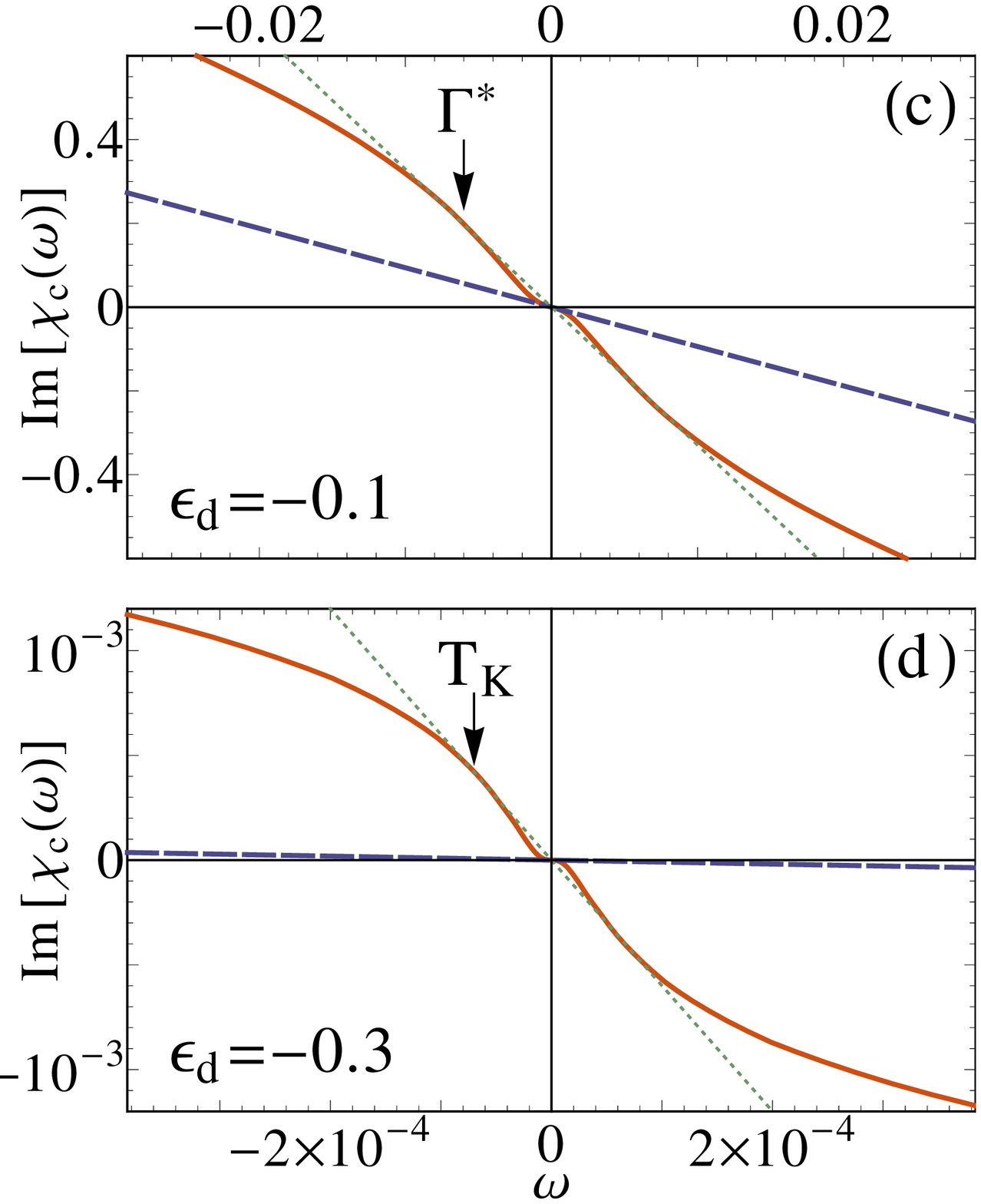}\hspace{8mm}%
  \includegraphics[width=35mm]{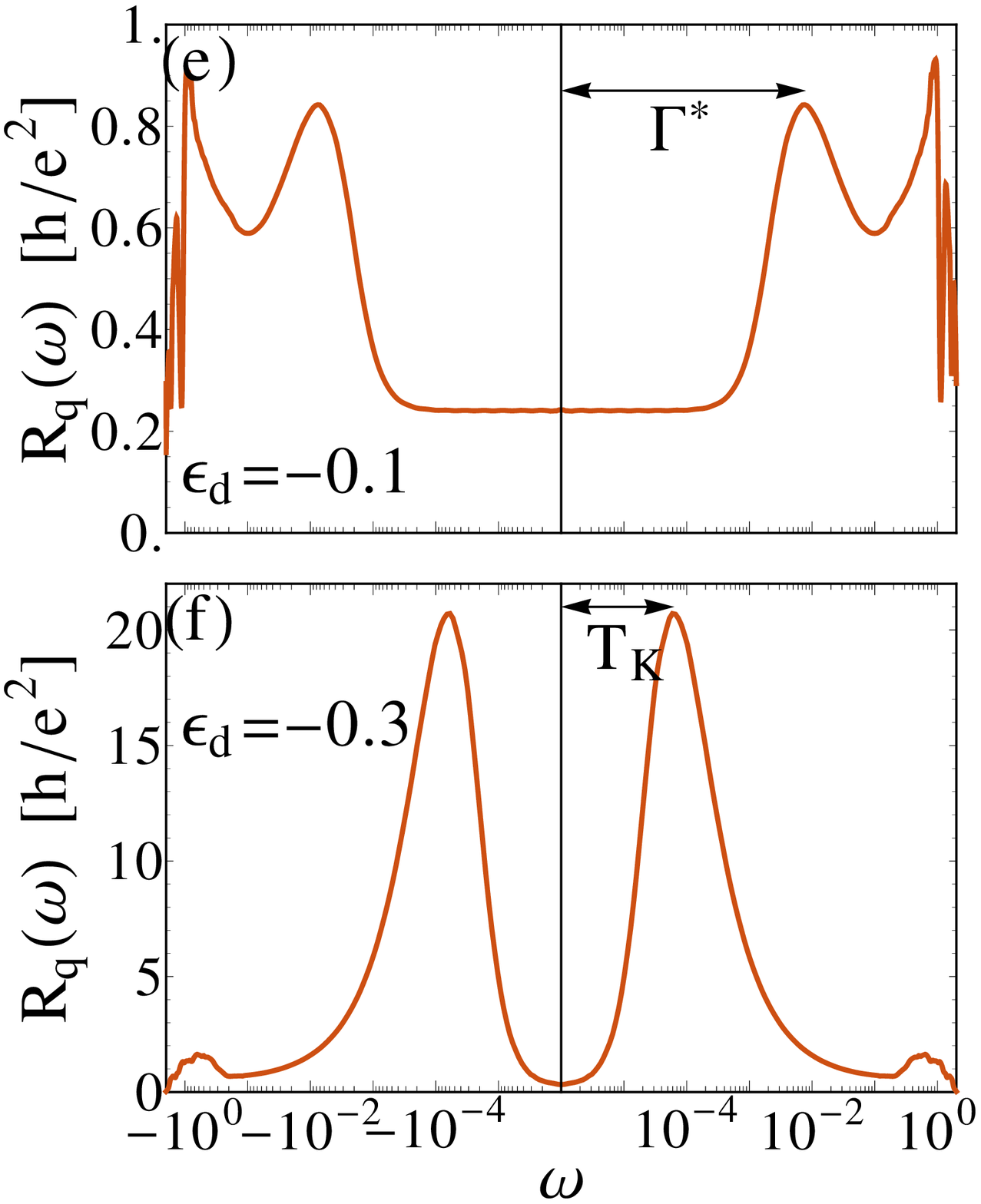}%
  \caption{(color online) (a) Zero-frequency limits of $R_q$ and $C_q$ versus
    $\epsilon_d$. (b) Typical spectral structure of the real and imaginary
    parts of the charge susceptibility $\chi_c$. (c,d) Enlarged views of
    $\im[\chi_c]$ near $\omega = 0$ for two values of $\epsilon_d$. The dashed
    (blue) and dotted (green) lines are tangent lines at infinitesimally small
    $\omega$ and at points (indicated by arrow) where the slope is maximal,
    respectively. (e, f) $R_q(\omega)$ in the logarithmic scale. Here $\Gamma =
    0.04$, and $E_{\rm C} = 0.5$.}
  \label{fig:2}
\end{figure}

\paragraph{No Zeeman splitting, $\Delta_Z=0$.---}
\Figref{fig:2}(a) shows our main results for $R_q$ and $C_q$ for the
spin-degenerate case.  First, the zero-frequency limit of the relaxation
resistance, $R_q(\omega\to0)$ is always close to the universal value $h/4e^2$,
regardless of values of $\epsilon_d$ and $E_{\rm C}$. This value can be
interpreted as the composite resistance of two parallel resistors of resistance
$h/2e^2$, the well-known universal resistance per channel
\cite{Buttiker:93a,Buttiker:93b,Nigg:06, Ringel:08}.  The NRG results show the
quantization of charge relaxation even in the Kondo regime where many-body
correlations are effective. The observed small deviations from the exact value
$h/4e^2$, persisting even in the noninteracting case, are attributed to the
finiteness of the contact bandwidth $D$, which introduces a frequency-dependent
real part into the dot self energy , $\re[\Sigma(\omega)] = - (\Gamma/2\pi)
\ln|(D-\hbar\omega)/(D+\hbar\omega)|$. Its presence slightly violates the
Fermi-liquid assumptions and, consequently, the KS relation is not exactly
fulfilled so that the universal value is not recovered: Since
$\re[\Sigma(\omega)]$ increases with $\omega$ in magnitude, the deviations are
larger as the resonant level becomes far from $\epsilon_F = 0$.  The universal
value can be restored by setting all the relevant energy scales to be much
smaller than $D$ \cite{Hewson:06}.  Second, the quantum capacitance, $C_q$
exhibits two remarkable considerations: \textit{(i)} at the degenerate points,
$\epsilon_d\sim\epsilon_F$ and $\epsilon_d + 2E_{\rm C} \sim \epsilon_F$, $C_q$
shows two pronounced peaks [see \figref{fig:2}(a)], which is consistent with
the known understanding that $C_q$ is proportional to the dot DOS $\rho_{\rm
  dot}(\epsilon_F)$ \cite{Buttiker:93a,Buttiker:93b,Buttiker:93c}, and
\textit{(ii) }$C_q$ remains quite small in the Kondo regime although the DOS at
the Kondo resonant level pinned at the Fermi level achieves its maximum
value. It implies that the Kondo resonant level, even though it can open a
tunneling channel, is not a real level which can hold real charges and cannot
contribute to the capacitance. Hence, in the presence of many-body correlations
$C_q$ is not, always, directly related to the DOS.

The frequency dependence of $R_q(\omega)$ and $C_q(\omega)$ is analyzed in
\figref{fig:2} where the the behavior of the real and imaginary part of the
charge susceptibility $\chi_c(\omega)$ is shown, see \figref{fig:2}(b) for
$\epsilon_d=-0.3$. The imaginary part of $\chi_c(\omega)$ reflects the coupling
between the ground state and $p$-$h$ excitations due to the dot-lead
hybridization. Since the spectral density of multiple $p$-$h$ excitations
increases with energy, $|\im[\chi_c]|$ would grow monotonically with
$|\omega|$. However, a finite $D$ puts an upper limit to the energy for $p$-$h$
excitations [$|\omega| \gtrsim \varO(D)$] resulting in the observed
nonmonotonic behavior for $\im[\chi_c]$.  Moreover, $\im[\chi_c]$ has two kinks
at $|\omega| = \min(|\epsilon_d|,|\epsilon_d+2E_{\rm C}|)$ since beyond this
frequency $p$-$h$ excitations accompanied with a charge excitation contributes
to $\im[\chi_c]$ as well.  An interesting structure appears in $\im[\chi_c]$
near $\omega = 0$, see \figsref{fig:2}(c) and (d) for two dot level positions,
$\epsilon_d=-0.1$ (corresponding to the fluctuating valence regime) and
$\epsilon_d=-0.3$ (Kondo regime).  Close to $\omega = 0$, $\im[\chi_c]$ depends
linearly with $\omega$, mainly due to single $p$-$h$ excitations [see tangent
(dashed) lines in \figsref{fig:2}(c) and (d)]. However, $\im[\chi_c]$ departs
from linearity when $\omega$ becomes of the order of the effective
hybridization $\Gamma^*$ ($=T_K$ in the Kondo regime). The renormalized
hybridization $\Gamma^*$($T_K$) is extracted from the width of the resonance
close to(at) $\epsilon_F$ in $\rho_{\rm dot}$.  Besides, we found that the
slope of $\im[\chi_c(\omega)]/\omega$ is the largest at $\hbar\omega =
\Gamma^*$, while the change in $\re[\chi_c(\omega)]$ is marginal. As a
consequence, $R_q(\omega)$ (see \eqnref{eq:Rq}) exhibits two side peaks at
$\hbar\omega = \pm\Gamma^*$ as shown in \figsref{fig:2}(e) and (f).  Notice
that in the Kondo regime $R_q(\omega)$ becomes much larger in order of
magnitude than the universal value, see \figref{fig:2}(f).  Remarkably, such
peak structure in $R_q(\omega)$ is absent in the noninteracting case. For a
noninteracting system, the analytical expression of $R_q(\omega)$ in the
wide-band limit is given by \cite{Jauho:94} $R_q^{\rm non}(\omega) = (h/e^2)
\{[G(\omega) + (\hbar\omega/\Gamma) F(\omega)]/[G(\omega)^2 + F(\omega)^2]\}$
with $G(\omega) = \ln\{[(\epsilon_d+\hbar\omega)^2+\Gamma^2]
[(\epsilon_d-\hbar\omega)^2+\Gamma^2]/[\epsilon_d^2 + \Gamma^2]\}$ and
$F(\omega) = 2\{\tan^{-1}[\Gamma/(\epsilon_d-\hbar\omega)] -
\tan^{-1}[\Gamma/(\epsilon_d+\hbar\omega)]\}$. $R_q^{\rm non}(\omega)$
increases monotonically with increasing $|\omega|$, and the only characteristic
energy scale is $\epsilon_d$.  Hence, the peaks seen in \figsref{fig:2}(e) and
(f) are a genuine many-body effect. In summary many-body correlations,
apparently having no impact on the zero-frequency value do affect $R_q(\omega)$
at finite frequencies, by forming a pronounced peak at $\hbar\omega =
\Gamma^*(T_K)$. The explanation for these observations will be given later.
\begin{figure}[!t]
  \centering
  \includegraphics[width=8.cm]{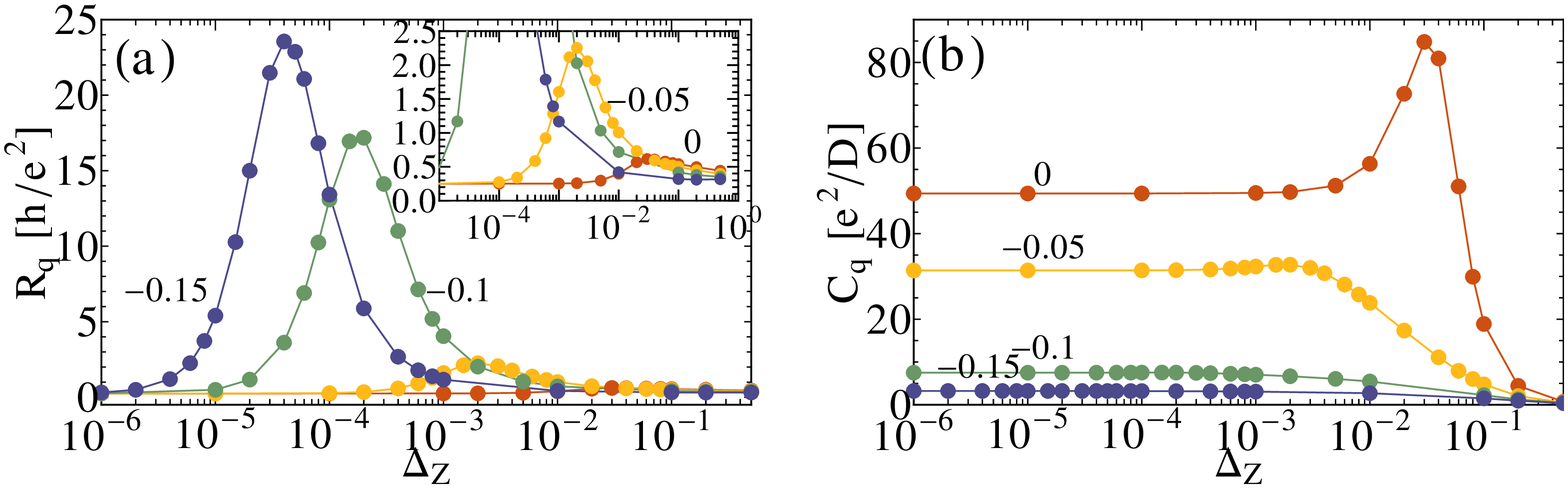}\\
  \includegraphics[width=4.2cm]{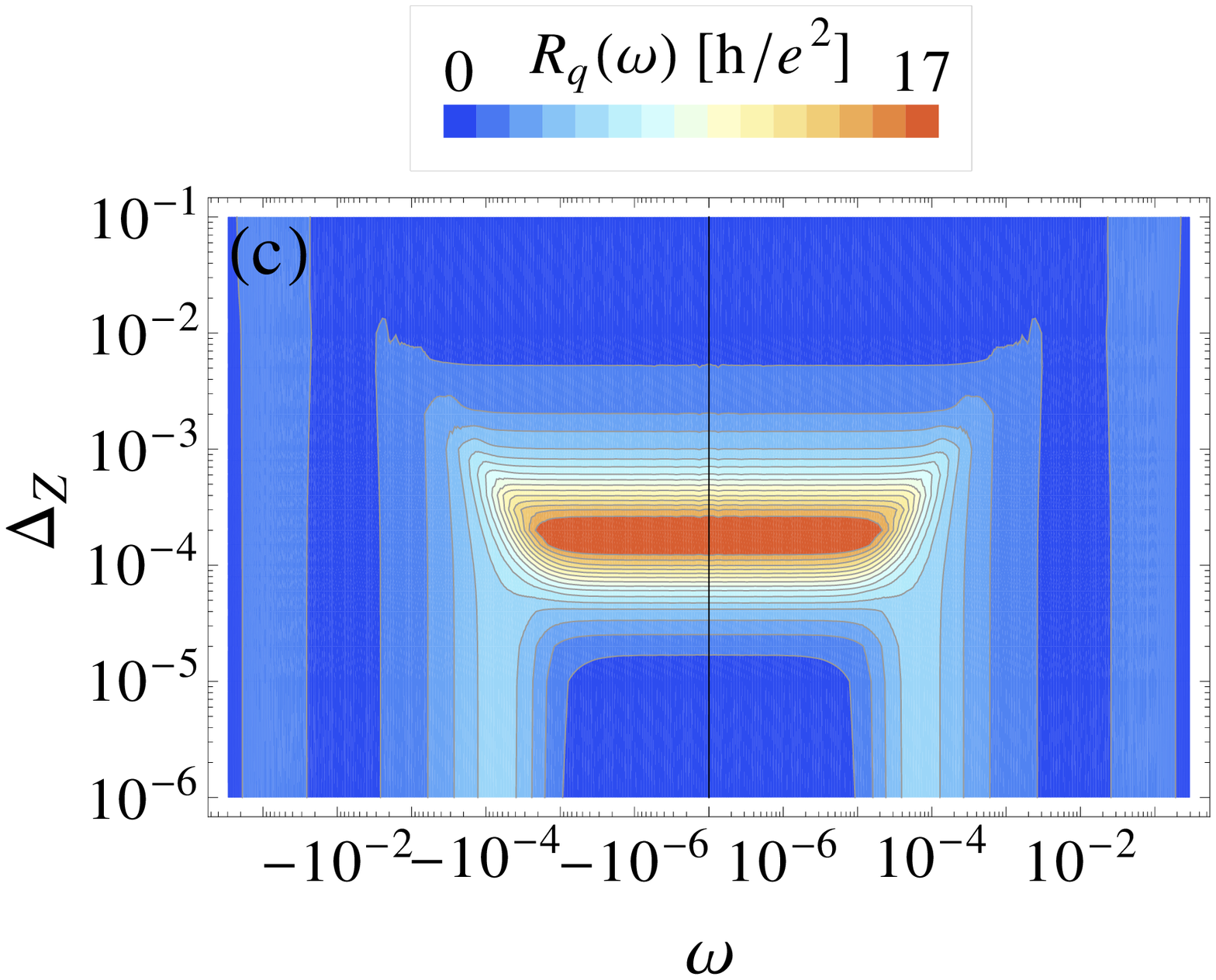}%
  \includegraphics[width=3.8cm]{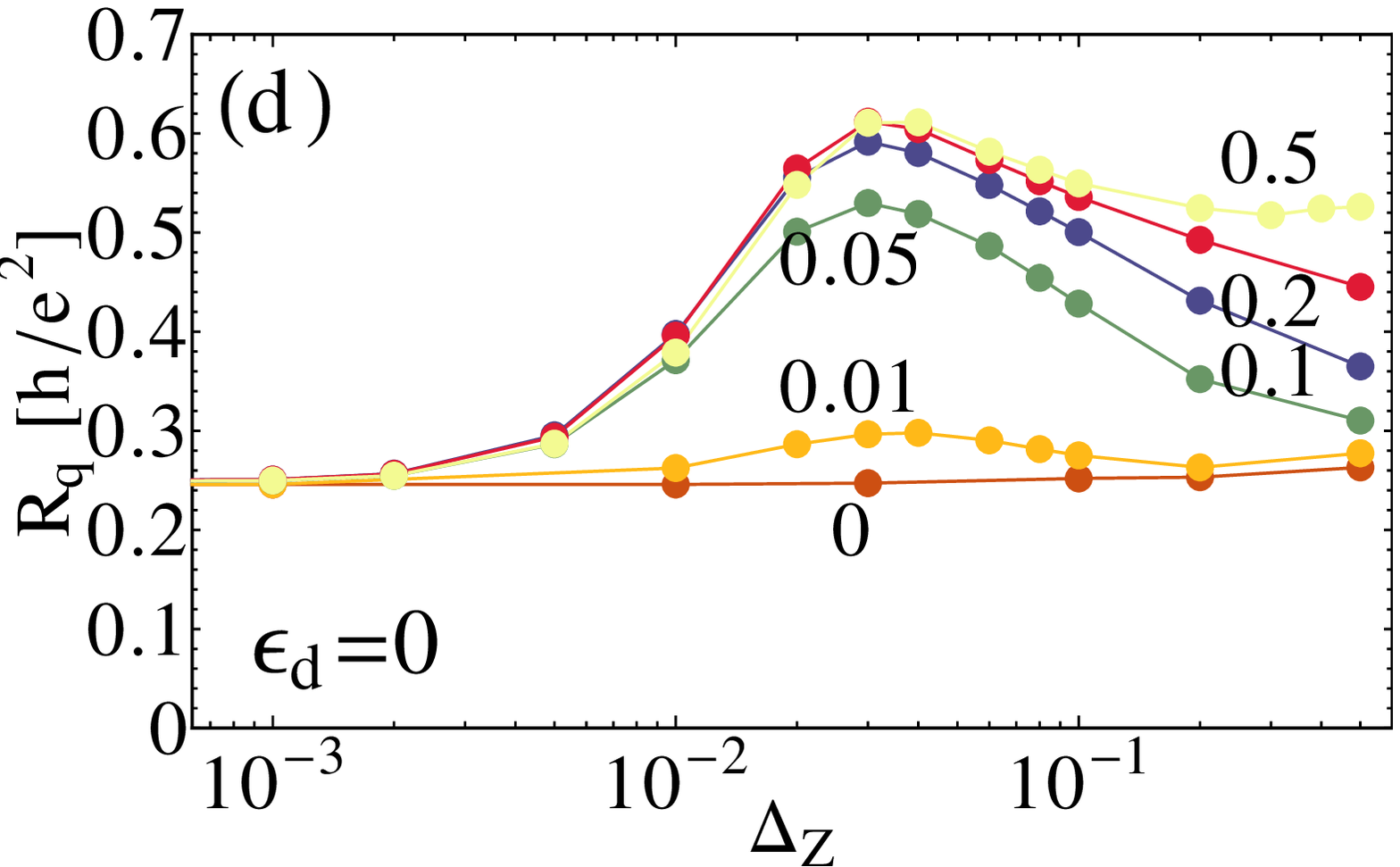}%
  \caption{(color online) (a,b) $R_q(\omega=0)$ and $C_q(\omega=0)$ versus
    $\Delta_Z$ for annotated values of $\epsilon_d$ with $E_{\rm C}=0.2$. The
    enlarged view of $R_q(\omega=0)$ for large $\Delta_Z$ is shown in the
    inset. (c) Contour plot of $R_q$ as a function of $\omega$ and $\Delta_Z$
    in the logarithmic scale in the Kondo regime ($\epsilon_d=-0.1$). (d)
    $R_q(\omega=0)$ versus $\Delta_Z$ for different values of $E_{\rm C}$ as
    annotated. Here $\Gamma = 0.02$ is used.}
  \label{fig:3}
\end{figure}

\paragraph{Finite Zeeman splitting, $\Delta_Z\neq 0$.---}
The spin-split case $(\Delta_Z>0)$ in the presence of external magnetic fields
is illustrated in \figref{fig:3}. Interestingly, $R_q(\omega\to0)$ versus
$\Delta_Z$ exhibits a peak structure reaching values much larger than the
quantized resistance in the spin-degenerate case: for example,
$R_q(\omega=0)|_{\rm max} \sim 100 \times h/4e^2$ for $\epsilon_d=-0.15$.
Furthermore, the peak is exactly located at $\Delta_Z = \Gamma^* (T_K)$ for the
fluctuating valence (Kondo) regime. The peak height increases as the the
effective hybridization decreases so it is the highest in the Kondo regime.
In the meanwhile, $C_q(\omega\to0)$ remains rather constant, except at the
resonant tunneling regime ($\epsilon_d\approx0$) where it displays a small
peak, see \figref{fig:3}(b).
The evolution of the spectral distribution of $R_q(\omega)$ with $\Delta_Z$ is
displayed in \figref{fig:3}(c) for $\epsilon_d=-0.3$ (Kondo regime). As
$\Delta_Z$ increases, the low-frequency part of $R_q(\omega)$ for
$|\hbar\omega| < k_BT_K$ keeps going up until $\Delta_Z$ reaches $k_BT_K$; the
side peaks are merged into the central peak. With increasing $\Delta_Z$
further, the central peak diminishes gradually and, eventually, together with
the side peaks located at $\hbar\omega = \pm k_BT_K$, disappear completely. We
have observed a similar transition of $R_q(\omega)$ with $\Delta_Z$ in the
resonant tunneling regime ($\epsilon_d = -0.05, 0$) except that the variation
of the central part is smaller.
Finally, \figref{fig:3}(d) compares the zero-frequency values of $R_q$ for
different values of the Coulomb interaction in the resonant tunneling
regime. In the noninteracting case, there is no peak at all, with
$R_q(\omega\to0)$ equal to $h/4e^2$. However, as soon as the charging energy
$2E_{\rm C}$ becomes comparable to $\Gamma^* \sim \Gamma$, a peak starts to
rise up and manifests itself for $2E_{\rm C} \gg \Gamma$. It implies that the
existence of the peak structure observed in \figref{fig:3}(d) definitely has
its origin in the Coulomb interaction.

\paragraph{Discussion.---}
Now we have two questions to be answered: \textit{(1)} How can Coulomb
interaction increase the relaxation resistance far beyond the universal value,
$h/4e^2$ and \textit{(2)} Why does it take place noticeably at $\hbar\omega =
\pm\Gamma^*(T_K)$ in the fluctuating valence (Kondo) regime for $\Delta_Z = 0$
or at $\hbar\omega = 0$ for $\Delta_Z = \Gamma^*(T_K)$? The charge relaxation
resistance is attributed to $p$-$h$ pair generation in the conduction band as
shown in \figref{fig:1}. Such processes are put in action when the dot-lead
tunneling is switched on. The tunneling in turn hybridizes dot and conduction
band electrons, resulting in lowering of the ground state by the effective
binding energy $\Gamma^*$ ($T_K$ in the Kondo regime). It means that the
$p$-$h$ generation starts when the energy supplied by the source is larger than
$\Gamma^*$. This argument explains the observed peak in $R_q(\omega)$ at
$\hbar\omega=\pm\Gamma^*$ in the absence of the Zeeman splitting. In the
presence of finite but small Zeeman splitting, the energy cost can be
compensated by the Zeeman splitting. The $p$-$h$ pair excitation states shown
in \figref{fig:1}(b) are now lowered by $\Delta_Z$, and when
$\Delta_Z\approx\Gamma^*$ they become almost degenerate with the ground state,
allowing $p$-$h$ pair generation with negligible energy cost. Hence,
$R_q(\omega)$ exhibits a single peak at $\hbar\omega=0$ when
$\Delta_Z=\Gamma^*$. This argument works solely when $\Delta_Z\lesssim\Gamma^*$
in which the ground state is not yet completely polarized and there exists a
finite coupling among spin-down dot states and spin-up dot states accompanying
with a $p$-$h$ pair generation in the reservoirs, see \figref{fig:1}. The
importance of the spin flip in the boosting of the relaxation resistance also
explains why $R_q(\omega)$ can reach higher values in the Kondo regime. The
Kondo ground state is built from spin fluctuations due to spin-flip scattering
among the localized dot electron and the delocalized electrons in the
reservoirs, thus spin-flip processes have large amplitudes in its
wavefunction. Hence the processes as shown in \figref{fig:1}(b) can happen more
frequently, leading to a large $R_q$.  Similarly, the answer for the first
question is now ready. The spectral weight for the charge correlation function
is proportional to $|\Braket{\alpha|\varN|\rm gs}|^2$, where $\alpha$ represent
the excited states.  In the second-order perturbation theory, this weight
corresponding to the processes in \figref{fig:1}(b) is given by
\begin{align}
  \label{eq:weight}
   |\Braket{\alpha|\varN|\rm gs}|^2
   =
   t^4
   \left|
     \frac{2}{E_\mu E_{\bar\mu}}
     {-}
     \sum_{\mu} \frac{\mu}{\Delta_Z}
     \left(
       \frac{1}{E_\mu} {+} \frac{1}{\epsilon_\mu}
     \right)
   \right|^2
 \end{align}
 in the $\omega\to0$ limit with $E_\mu=2E_{\rm C}+\epsilon_\mu$. Interestingly,
 this weight vanishes for $E_{\rm C} = 0$ for any value of $\Delta_Z$. Thus,
 for the noninteracting case there exists no $p$-$h$ pair generation process
 accompanying spin flip in the dot, and no boosting of the relaxation
 resistance can happen. For finite values of $E_{\rm C}$, the weight is finite
 [see \eqnref{eq:weight}] and for $E_{\rm C}\to\infty$, it becomes
 $t^4/(\epsilon_\up\epsilon_\down)^2$. This value can be substantial depending
 on the level position. Note that this analysis is not correct quantitatively
 because high-order events should be considerably involved in the observed
 phenomena. The observed boosting of $R_q$ at $\Delta_Z \sim \Gamma^*$
 indicates that the perturbation in the dot-lead tunneling or $\Gamma$ is
 risky. A more general theoretical analysis that treats $\Delta_Z$ and $\Gamma$
 on equal footing could provide more quantitatively reliable
 interpretation. Besides, this perturbative analysis does not work in the Kondo
 regime where the strong dot-lead coupling is important. One may want to study
 the Kondo regime by an effective single-particle Hamiltonian with a dot level
 at the Fermi energy with the effective hybridization $T_K$. However, this
 picture is only suitable in the Fermi-liquid regime in which $p$-$h$
 excitations accomplished by spin-flip events in the dot are not
 allowed. Besides, this effective model predicts an enhanced mesoscopic
 capacitance $C_q(\omega\to0)$ due to the presence of the resonant level at the
 Fermi level. As noted before, however, the Kondo resonant level cannot
 contribute to the charging of real charges.

\paragraph{Conclusion.---}
In closing, we have investigated the dynamics of a many-body quantum
capacitor. Using the relation of charge relaxation resistance and quantum
capacitance with the charge susceptibility, we find that in the deep Kondo
regime the KS rule ensures a quantized $R_q(\omega=0)=h/4e^2$. Besides we show
that the interpretation of $C_q$ in terms of the DOS fails when many-body
effects are present. Here, $C_q(\omega=0)$ becomes very small even when
$\rho_{\rm dot}(\epsilon_F)$ becomes large due to a completely frozen charge
dynamics in the presence of Kondo correlations.  Finally, we find that
$R_q(\omega)$ is built by dot-lead tunneling events connecting $p$-$h$
excitations in the reservoirs with spin-flip processes in the dot . This
interpretation explains our results for $R_q(\Delta_Z=0,\omega)$ showing peaks
at $\hbar\omega=\pm\Gamma^*(T_K)$ and for $R(\Delta_Z=\Gamma^*(T_K),\omega)$
exhibiting a peak at $\hbar\omega=0$.

\begin{acknowledgments}
  M.L. was supported by the NRF Grants (2009-0069554 and
  2010-0015416). R. L. was supported by MEC-Spain (Grant
  No. FIS2008-00781). T.M. and T.J. acknowledge support from ANR 2010 BLANC
  0412 02. M.-S.C. was supported by the NRF Grants (2009-0080453 and
  2010-0025880).
\end{acknowledgments}


\begin{thebibliography}{99}
\bibitem{nrgnote} For better accuracy, we adopt Hofstetter's algorithm
  \cite{Hofstetter:00} accompanied with the improved $z$-averaging method
  \cite{NRGzavg}.
\end{thebibliography}
\end{document}